\documentstyle[floats,aps,psfig,twocolumn]{revtex}
\begin{document}

\draft
\twocolumn[\hsize\textwidth\columnwidth\hsize\csname
@twocolumnfalse\endcsname

\title{Phase transitions in optimal unsupervised learning}

\author{Arnaud Buhot \and Mirta B. Gordon\cite{cnrs}}
\address{D\'epartement de Recherche Fondamentale sur la Mati\`ere
Condens\'ee,\\ CEA/Grenoble, 17 rue des Martyrs, 38054 Grenoble Cedex 9,
France}

\date{\today}

\maketitle

\begin{center}
\begin{abstract}
\parbox{14cm}{We determine the optimal performance of learning the
orientation of the symmetry axis of a set of $P= \alpha N$ points
that are uniformly distributed in all the directions but one on the
$N$-dimensional space. The components along the symmetry breaking
direction, of unitary vector ${\bf B}$, are sampled from a mixture of
two gaussians of variable separation and width. The typical optimal
performance is measured through the overlap $R_{\rm opt}={\bf B}
\cdot {\bf J}^*$ where ${\bf J}^*$ is the optimal guess of the
symmetry breaking direction. Within this general scenario, the
learning curves $R_{\rm opt}(\alpha)$ may present first order
transitions if the clusters are narrow enough. Close to these
transitions, high performance states can be obtained through the
minimization of the corresponding optimal potential, although these
solutions are metastable, and therefore not learnable, within the
usual bayesian scenario.}
\end{abstract}
\end{center}

]

\section{Introduction}

In this paper we address a very general problem in the statistical 
analysis of large amounts of data points, also called {\it examples}, 
{\it patterns} or {\it training set}, namely the one of discovering 
the structure underlying the data set. Whether this determination is 
possible or not depends on the assumptions one is willing to 
accept~\cite{Duda}. Several algorithms allowing to detect structure 
in a set of points exist. Among them, principal component analysis 
finds the directions of higher variance, projection pursuit 
methods~\cite{Ripley} seek directions in input space onto which the 
projections of the data maximize some measure of departure from 
normality, whereas self-organizing clustering 
procedures~\cite{Kohonen} allow to determine prototype vectors 
representative of clouds of data. The parametric approach assumes 
that the structure of the probability density function the patterns 
have been sampled from is known. Only its parameters have to be 
determined given the examples. A frequent guess is that the 
probability density is either gaussian, or a mixture of gaussians. 
The process of determining the corresponding parameters is called 
{\it unsupervised learning}, because we are not given any additional 
information about the data, in contrast with {\it supervised 
learning} in which each training example is labelled. 

It has recently been shown that finding the principal component of a 
set of examples, clustering data with a mixture of gaussians, and 
learning pattern classification from examples with neural networks 
may be casted as particular cases of unsupervised 
learning~\cite{RV1}. In all these problems, the examples are drawn 
from a probability density function ({\it pdf}) with axial symmetry, 
and the symmetry-breaking direction has to be determined given the 
training set. As this direction may be found through the minimization 
of a cost function, the properties of unsupervised learning may be 
analyzed with statistical mechanics.  This approach allows to 
establish the properties of the typical solution, determined in the 
thermodynamic limit, {\it i.e.} the space dimension $N \rightarrow 
+\infty$, the number of examples $P \rightarrow +\infty$, with the 
fraction of examples $\alpha = P/N$ constant. 

Besides these general results, the statistical mechanics framework 
allows to deduce the expression of an {\it optimal cost 
function}~\cite{KC1,BTG1,VR1}, whose minimum is the best solution 
that may be expected to be learnt given the data. The optimal cost 
function depends on the functional structure of the {\it pdf} the 
examples are sampled from, and on the fraction $\alpha$ of available 
examples. Its main interest is that it allows to deduce the upper 
bound for the typical performance that may be expected from any 
learning algorithm. On the other hand, Bayes' formula of statistical 
inference allows to determine the probability of the symmetry 
breaking direction given the training set. Sampling the direction 
with Bayes probability is called Gibbs learning~\cite{WN1}. The 
average of the solutions obtained through Gibbs learning, weighted 
with the corresponding probability, is called {\it bayesian} 
solution. It is widely believed that the bayesian solution is 
optimal. Moreover, this has been so in all the scenarios considered 
so far. 

In the present paper, we consider a very general two-cluster 
scenario, which contains results already reported as particular 
cases. In fact, two different situations, in which the pattern 
distribution is a gaussian of zero mean and unit variance in all the 
directions but one, have been considered so far: a gaussian 
scenario~\cite{RVB1} and a two-cluster scenario~\cite{BM1,BM2,WN1}. 
In the former, the components of the examples parallel to the 
symmetry-breaking direction are sampled from a single gaussian. In 
the latter these components are drawn from a mixture of two 
gaussians, each one having unit variance. The learning process has to 
detect differences between the {\it pdf} along the symmetry-breaking 
direction and the distributions in the orthogonal directions. Several 
{\it ad hoc} cost functions allowing to determine the 
symmetry-breaking direction have been analyzed for both scenarios. 
Typically, if the {\it pdf} has a non-zero mean value in the 
symmetry-breaking direction, learning is "easy": the quality 
of the solution increases monotonically with the fraction 
$\alpha$ of examples, starting at $\alpha=0$. In contrast, 
if the {\it pdf} has  zero mean, the deviations of the 
{\it pdf} along the symmetry breaking direction from the 
{\it pdf} in the orthogonal directions depend on the second 
and higher moments. In this case, a phenomenon called {\it 
retarded learning}~\cite{WN1} appears: learning the 
symmetry-breaking direction becomes impossible when 
the fraction of examples falls below a critical value $\alpha_c$.

Since we have considered the case of clusters of variable width, we
could determine the entire phase diagram of the two-cluster scenario.
Several new learning phases appear, depending on the mean and the
variance of the clusters. In particular, if the second moment of the 
individual clusters is smaller than the second moment of the {\it pdf} 
in the orthogonal directions, first order transitions from low to high
performance learning may occur as a function of $\alpha$. Close to
these, high performance metastable states exist above the stable
states of Gibbs learning, in the thermodynamic limit. One of the
most striking results of this paper is that these high performance
metastable states can indeed be learnt through the minimization 
of an optimal $\alpha$-dependent potential, although they cannot 
be obtained through bayesian learning. 

Our results have been obtained within the replica approach with the
replica symmetry hypothesis. We show below that this assumption is
equivalent to the more intuitive requirement that the optimal
learning curves $R_{\rm opt}(\alpha)$ are increasing functions of the
fraction of examples $\alpha$. To our knowledge, this fact has not
been noticed before.

The paper is organized as follows: a short presentation of the
problem and the replica calculation are given in section
\ref{Gen_Frame}. In section \ref{Opt_Pot} we deduce the optimal cost
functions within the replica symmetry hypothesis, as well as the
condition of replica symmetry stability. In section
\ref{Double_Gauss} we deduce and discuss the optimal learning
curves for the general two-cluster scenario. The typical 
properties of the optimal cost functions in the complete 
range of $\alpha$, presented in section \ref{B_vs_Opt}, 
show that bayesian learning may not be optimal. Finally, the 
complete phase diagram is described in section \ref{sec:phase_diag}, 
as a function of the two clusters' parameters.

\section{General framework and replica calculation.}
\label{Gen_Frame}

We consider the general case of $N$-dimensional vectors ${\bbox
\xi}$, the patterns or examples of the training set, drawn from an
axially symmetric probability density $P^*({\bbox \xi} \, | {\bf B})$
of the form:

\begin{equation}
\label{eq:proba}
P^*({\bbox \xi} \, | {\bf B}) \equiv  \frac{1}{(2 \pi)^{N/2}} \,
\exp \left\{ {- \frac{{\bbox \xi} \cdot {\bbox \xi}}{2}}
- V^*( \lambda)\right\},
\end{equation}

\noindent where ${\bf B}$ is a unitary vector in the
symmetry-breaking direction, {\it i.e.} ${\bf B} \cdot {\bf B} = 1$
(notice that this is {\it not} the usual convention), and $\lambda
\equiv {\bbox \xi} \cdot {\bf B} = \sum_i \xi_i B_i$. According to
(\ref{eq:proba}), the patterns have normal distributions {\it i.e.}
$P(x) = \exp (-x^2/2) / \sqrt{2 \pi}$ onto the $N-1$ directions
orthogonal to ${\bf B}$. The distribution (\ref{eq:proba}) in the
symmetry-breaking direction is

\begin{equation}
\label{eq:broken_proba}
P^* ( \lambda ) = \frac{1}{ \sqrt{2 \pi} } \exp \left\{ - \frac{\lambda
^2}{2} - V^* ( \lambda ) \right\}.
\end{equation}

\noindent Thus, $V^* ( \lambda )$ introduces a modulation parallel to
${\bf B}$; if $V^* = 0$ the patterns' distribution is normal in all
the directions. Normalization of $P^*$ requires:

\begin{equation}
\int_{-\infty}^{+\infty} D \lambda\, \exp \left[ - V^* ( \lambda )
\right] = 1
\end{equation}

\noindent where $D \lambda = \exp (- \lambda ^2/2) d \lambda /
\sqrt{2 \pi}$. The different moments $\langle \lambda ^{n} \rangle$
of (\ref{eq:broken_proba}) are:

\begin{equation}
\label{eq:moments}
\langle \lambda ^{n} \rangle \equiv \int ({\bbox \xi} \cdot {\bf B})^n P^*({\bbox
\xi} \, | {\bf B}) d{\bbox \xi} = \int_{-\infty}^{+\infty} \lambda^n P^* (\lambda )\, d \lambda.
\end{equation}
 
Several examples of functions $V^*$ have been treated in the
litterature so far~\cite{RV1,VR1,WN1,RVB1,BM1,BM2}. In the particular
case of supervised learning of a linearly separable classification
task by a single unit neural network, the symmetry-breaking direction
${\bf B}$ is the {\it teacher's} vector, orthogonal to the hyperplane
separating the classes. The class of pattern ${\bbox \xi}$ is $\tau
\equiv sign({\bf B} \cdot {\bbox \xi})$. The corresponding {\it pdf}
is $P^* (\tau \lambda) = 2 \, \Theta (\tau \lambda) \, \exp
(-\lambda^2/2)/\sqrt{2 \pi}$, {\it i.e.} $V^* (\lambda) = - \ln 2 $
for $\tau \lambda > 0$ and $ + \infty $ for $\tau \lambda < 0$.

In the following, we concentrate on the problem of unsupervised
learning. We are given a {\it training} set ${\cal L}_{\alpha} = \{
{\bbox \xi}^{\mu} \}_{\mu = 1, \ldots,P}$ of $P=\alpha N$ vectors
sampled independently with probability density $P^* ({\bbox \xi}
\, | {\bf B})$. We have to {\it learn} the unknown symmetry-breaking
direction ${\bf B}$ from the examples knowing the functional
dependence of $P^*$ on ${\bf B}$. Using Bayes' rule of inference, the
probability of a direction ${\bf J}$ (with ${\bf J} \cdot {\bf J}=1$)
given the data is:

\begin{equation}
\label{eq:Gibbs}
P({\bf J} | {\cal L}_{\alpha} ) = \frac{1}{\cal Z}
\prod_\mu \exp \left\{-{\bbox \xi}^\mu \cdot {\bbox \xi}^\mu/2 -
V^*( {\bbox \xi}^\mu \cdot {\bf J} )\right\} P_0({\bf J}),
\end{equation}
 
\noindent where $P_0({\bf J}) = \delta ({\bf J} \cdot {\bf J}-1)$ is
the assumed prior probability and ${\cal Z}=\int d {\bf J} \prod_\mu
\exp \left\{-{\bbox \xi}^\mu \cdot {\bbox \xi}^\mu/2 - V^*( {\bbox
\xi}^\mu \cdot {\bf J} )\right\} P_0({\bf J})$ is the probability of
the training set. By analogy with supervised learning, sampling the
direction with probability (\ref{eq:Gibbs}) is called {\it Gibbs
learning}~\cite{WN1}.

We consider learning procedures where the direction ${\bf J}$ is
found through the minimisation of a cost function or energy $E \left(
{\bf J}; {\cal L}_{\alpha}\right)$. As the patterns are independently
drawn, this energy is an additive function of the examples. The
contribution of each pattern ${\bbox \xi}^{\mu}$ to $E$ is given by a
{\it potential} $V$ that depends on the direction ${\bf J}$ and on
${\bbox \xi}^{\mu}$ through the projection  (called {\it local
field}) $\gamma^{\mu} = {\bf J} \cdot {\bbox \xi}^{\mu}$:

\begin{equation}
\label{eq:energy}
E \left( {\bf J};{\cal L}_{\alpha} \right) = \sum_{\mu = 1}^P V \left( \gamma^{\mu}
\right).
\end{equation}

\noindent As the training set only carries partial information on the
symmetry-breaking direction ${\bf B}$, the direction ${\bf J}$
determined by the minimization of (\ref{eq:energy}) will generally
differ from ${\bf B}$. The quality of a solution ${\bf J}$ may be
caracterized by the overlap $R = {\bf B} \cdot {\bf J}$. If $R=0$,
${\bf J}$ does not give any information about the symmetry-breaking
direction. Conversely, if $R=1$ the symmetry-breaking direction is
perfectly determined.

The statistical mechanics approach allows to calculate the expected
overlap $R(\alpha)$ for any general distribution $V^*$ and any
general potential $V$, in the thermodynamic limit $N,P \rightarrow +
\infty$ with $\alpha \equiv P/N$ finite. In this limit, we expect that the energy is
self-averaging: its distribution is a delta peak centered at its
expectation value independently of the particular realization of the
training patterns. Given the modulation $V^*$, different values of
$R$ may be reached, depending on the potential used for learning. In
the following, we sketch the main lines that allow to derive the
typical value of $R$ corresponding to a general potential $V$.

The free energy $F$ corresponding to the energy (\ref{eq:energy})
with a given potential $V(\gamma)$ is

\begin{equation}
\label{eq:free_en}
F ( \beta, N, {\cal L}_{\alpha} ) = -
\frac{1}{\beta} \ln Z( \beta, N, {\cal L}_{\alpha}),
\end{equation}

\noindent where $\beta$ is the inverse temperature and $Z$ the
partition function:

\begin{equation}
Z( \beta, N, {\cal L}_{\alpha} ) = \int d
{\bf J} \, \exp \left\{ - \beta E({\bf J};{\cal L}_{\alpha}) \right\} \delta ( {\bf J}^2 - 1).
\end{equation}

\noindent As mentioned before, in the thermodynamic limit the free
energy is self-averaging, {\it i.e.}:

\begin{equation}
\label{eq:self_aver}
\lim_{N \rightarrow +\infty} \frac{1}{N} \, F(\beta,N,{\cal L}_{\alpha} ) = \lim_{N \rightarrow + \infty} \frac{1}{N} \,
\overline{ F(\beta,N,{\cal L}_{\alpha} ) }
\end{equation}

\noindent where $\overline{ ( \ldots ) }$ stands for the average over
all the possible training sets. The average in the right hand side of
eq. (\ref{eq:self_aver}) is calculated using the replica method:

\begin{equation}
\overline{ \ln Z } = \lim_{n \rightarrow 0} \frac{1}{n} \ln
\overline{ Z^n }.
\end{equation}

\noindent which reduces the problem of averaging $\ln Z$ to the one
of averaging the partition function of $n$ {\it replicas} of the
original system, and taking the limit $n \rightarrow 0$. The
properties of the minimum of the cost function are those of the zero
temperature limit ($\beta \rightarrow +\infty$) of the free energy.
In the case of differentiable potentials $V$, the integrals are
dominated by the saddle point, and the zero temperature free energy
writes~\cite{RV1}:

\begin{eqnarray}
\label{eq:f(R,c)}
f(R,c) & = & \lim_{ \beta \rightarrow + \infty } \lim_{ N \rightarrow
+ \infty } \frac{1}{N} \overline{ F(\beta,N,{\cal L}_{\alpha} ) }\\ 
\nonumber & = & - \frac{1}{2c} \left\{ 1 - R^2 - 2 \alpha \, \int Dt \, W(t;c) \right.\\
&   & \left. \times \int Dz \, \exp \left[-V^*(\lambda) \right] \right\},
\nonumber
\end{eqnarray}

\noindent where 

\begin{equation}
\label{eq:lambda}
\lambda \equiv z \sqrt{1-R^2} + Rt.
\end{equation}

\noindent In (\ref{eq:f(R,c)}), $R$ is the overlap between the
symmetry-breaking direction ${\bf B}$ and a minimum ${\bf J}$ of the
cost function (\ref{eq:energy}); $c = \lim_{\beta \rightarrow
+\infty} \beta (1 - q)$ where $q$ is the overlap between minima of
the cost function (\ref{eq:energy}) for two different replicas, and

\begin{equation}
\label{eq:saddle_p}
W(t;c) = {\rm min}_{\, \gamma} \left[ cV(\gamma) + (\gamma - t)^2/2
\right],
\end{equation}

\noindent is the saddle point equation. The extremum conditions of
the free energy (\ref{eq:f(R,c)}) with respect to $R$ and $c$,
$\partial f/\partial R=\partial f/\partial c=0$, give the following
equations for $R$ and $c$:

\begin{mathletters}
\label{eq:R1_R2}
\begin{eqnarray} 
\nonumber 1 - R^{\, 2} & =  \alpha & \int_{-\infty}^{+\infty} Dt \, [\gamma(t;c) - t]^2 \\
\label{eq:R1}  &  & \times \int_{-\infty}^{+\infty} Dz \, \exp \left[- V^*(\lambda) \right], \\
\nonumber R\,\sqrt{1-R{\,^2}} & =  \alpha & \int_{-\infty}^{+\infty} Dt \, [\gamma(t;c) - t] \\
\label{eq:R2} &  & \times \int_{-\infty}^{+\infty} Dz \, z\, \exp \left[- V^*(\lambda) \right],
\end{eqnarray}
\end{mathletters}

\noindent where $\lambda$ is defined in (\ref{eq:lambda}) and
$\gamma(t;c)$ is the solution that minimizes (\ref{eq:saddle_p}).
Introduction of (\ref{eq:R1_R2}) into (\ref{eq:f(R,c)}) gives the
free energy at zero temperature:

\begin{equation}
\label{eq:f(T=0)}
f(R,c) =  \alpha \, \int Dt \, V(\gamma(t;c)) \int Dz \, \exp
\left[-V^*(\lambda) \right].
\end{equation}

\noindent If the potential $V(\gamma)$ is not convex, eq.
(\ref{eq:R1_R2}) may have more than one solution. In that case, the
one minimizing (\ref{eq:f(T=0)}) with respect to $R$ should be kept.

These results were obtained under the assumption of replica symmetry.
A necessary condition for the replica symmetry hypothesis to be
satisfied is:

\begin{equation}
\label{eq:rep_sym}
\alpha \int_{-\infty}^{+\infty} Dt \, \left[\gamma'(t;c) -
1\right]^2 \int_{-\infty}^{+\infty} Dz \, \exp \left[ -
V^*(\lambda) \right] < 1,
\end{equation}

\noindent with $\gamma'(t;c) \equiv \partial \gamma / \partial t$.

\section{Optimal potential and replica symmetry stability condition.}
\label{Opt_Pot}

Given any modulation $V^*$, the typical overlap $R$ obtained through
the minimization of a differentiable potential $V$ may be determined
as a function of $\alpha$ by solving equations (\ref{eq:R1_R2}). The
result is consistent if condition (\ref{eq:rep_sym}) is verified. In
this section, we are interested on the {\it best} performances that
may be expected. Recently, a general expression for the optimal
potential allowing to find the solution with maximum overlap $R_{\rm
opt}$ has been deduced~\cite{RV1}. This {\it optimal potential}
$V_{\rm opt}$ depends implicitly on $\alpha$ through $R_{\rm
opt}(\alpha)$, and on the probability distribution $P^*$ through the
modulation $V^*$. It was obtained under the assumption of replica
symmetry, which has been shown to be correct for the particular cases
investigated so far. In fact, the stability condition of replica
symmetry for optimal learning is verified whenever the slope of the
learning curves is positive, as will be shown below. For the sake of
completness, we first describe an alternative derivation of the
optimal potential. Following the same lines we used for supervised
learning~\cite{BTG1}, $V_{\rm opt}$ is determined through a functional
maximization of $R$, given by eq. (\ref{eq:R1_R2}), with respect to
$V$ at constant $\alpha$. As discussed in~\cite{BTG1}, the parameter 
$c$ sets the energy units and may be arbitrarily chosen. We used 
$c=1$ throughout, without any lack of generality. After a straightforward calculation we obtain that the optimal overlap $R_{\rm opt}$ is given 
by the inversion of:

\begin{equation}
\label{eq:alpha(R)}
\alpha (R_{\rm opt}) = R_{\rm opt}^{\, 2} \left\{\int_{-\infty}^{+\infty} Dt \, \frac{\left[ \int Dz \, z \exp \left(- V^*(\lambda) \right) \right]^2} {\int Dz \exp \left(- V^*(\lambda) \right)} 
\right\}^{-1},
\end{equation}

\noindent  where $\lambda$, given by (\ref{eq:lambda}), writes
$\lambda \equiv z \sqrt{1-R^2_{\rm opt}} \, + R_{\rm opt} \, t$.
Notice that eq. (\ref{eq:alpha(R)}) may be not invertible, {\it
i.e.}, $R_{\rm opt}(\alpha)$ may be multivalued. In this case, the
correct solution has to be selected. 

$V_{\rm opt}$ is determined through the integration of:

\begin{equation}
\label{eq:Vopt}
V'_{\rm opt}(\gamma_{\rm opt}(t)) = \frac{1-R^2_{\rm opt}}{R^2_{\rm opt}}
\frac{d \,}{dt} \left[ \ln \int_{-\infty}^{+\infty} Dz \exp \left( -V^*(\lambda) \right) \right],
\end{equation}

\noindent where the argument of $V'_{\rm opt}$ is given by the
saddle-point equation (\ref{eq:saddle_p}) with $c=1$, {\it i.e}:

\begin{equation}
\label{eq:gamma(t)}
\gamma_{\rm opt}(t) = t-V'_{\rm opt}(\gamma_{\rm opt}(t)).
\end{equation}

\noindent Since $R$ is parametrized by $\alpha$, the cost function 
leading to optimal performance is different for different training 
set sizes. 

Eq. (\ref{eq:alpha(R)}) and (\ref{eq:Vopt}) were previously
derived by Van den Broeck and Reiman~\cite{VR1}, who showed that the
typical overlap $R_{\rm b}$ of bayesian learning satisfies the same
equation (\ref{eq:alpha(R)}) as $R_{\rm opt}$. However, this only
guarantees that bayesian learning is optimal if eq. (\ref{eq:alpha(R)}) 
is invertible. In that case its unique solution is $R_{\rm b}=R_{\rm opt}$. Otherwise, as is discussed in the example of section \ref{Double_Gauss},
solutions with $R_{\rm opt} > R_{\rm b}$ may exist.

The results derived so far are valid under the replica symmetry
hypothesis, and must thus satisfy (\ref{eq:rep_sym}). Taking
(\ref{eq:alpha(R)}) and (\ref{eq:gamma(t)}) into account, a
cumbersome but straightforward calculation gives:

\begin{eqnarray}
\nonumber 1 & - & \alpha \int_{-\infty}^{+\infty} Dt \, \left[\gamma'(t;c) -
1\right]^2 \int_{-\infty}^{+\infty} Dz \, \exp \left[ -
V^*(\lambda) \right]\\
\label{eq:rsc_equiv} & = & \frac{R_{\rm opt}^2(1-R_{\rm opt}^2)}{\alpha} \, 
\frac{d \alpha(R_{\rm opt})}{d R_{\rm opt}^2}
\end{eqnarray}

\noindent Therefore, in the case of optimal learning, the necessary
condition of replica symmetry stability (\ref{eq:rep_sym}) {\it is
equivalent} to the natural requirement that the learning curve
$R_{\rm opt}(\alpha)$ is an increasing function of the fraction of
examples $\alpha$ for $R_{\rm opt} \neq 0,1$. This relation, which
does not seem to have been noticed before, is independent of the
distribution (\ref{eq:proba}) the data set is sampled from. 

In the cases where the analytic function $\alpha \, (R_{\rm opt})$ given
by (\ref{eq:alpha(R)}) is not invertible, only the branches
with positive slope have to be considered, as they trivially satisfy the
replica symmetry condition. Examples of such a behaviour are shown
in next section.

Hence, given any modulating function $V^*$ sufficently derivable, as
far as $R_{\rm opt} \neq  0,1$ there exists an optimal potential
$V_{\rm opt}(\gamma)$, consistent with the assumptions of the replica
calculation, which depends implicitly on $\alpha$ through $R_{\rm
opt}(\alpha)$, and on $V^*$. The minimum ${\bf J}^*$ of the
corresponding energy (\ref{eq:energy}) maximizes the overlap $R$
between ${\bf J}^*$ and the symmetry-breaking direction ${\bf B}$.

The development of $\alpha \, (R_{\rm opt})$ for small $R_{\rm opt}$
shows that $R_{\rm opt}>0$ for all $\alpha > 0$ if and only if
$\langle \lambda \rangle \, \neq 0$. In that case, for $\alpha \ll
1$, $R_{\rm opt} \approx \, \langle \lambda \rangle \sqrt{\alpha}$,
like with Hebb's learning rule~\cite{RV1}. If $\langle \lambda \rangle\, 
= 0$, two different behaviours may arise: either a continuous 
transition from $R_{\rm opt}=0$ to $R_{\rm opt} \sim \sqrt{\alpha - 
\alpha_c}$ occurs at $\alpha_c \equiv (1 - \langle \lambda^2 
\rangle)^{-2}$, or the overlap jumps from $R_{\rm opt} = 0$ to 
$R_{\rm opt} > 0$ through a first order transition at $\alpha_1 
\leq \alpha_c$. In particular, if $\langle \lambda^2 \rangle \, = 
1$, only a discontinuous transition may occur since $\alpha_c = 
+ \infty$. Discontinuities between two finite values of $R_{\rm opt}$ 
also may arise for $\alpha > \alpha_c$. All these phase transitions appear 
in the two-cluster scenario that we analyze in next section.

\section{A case study: two-cluster distributions.}
\label{Double_Gauss}

Consider the general two gaussian-clusters scenario, in which the
modulation along the symmetry breaking direction
(\ref{eq:broken_proba}) is:

\begin{equation}
\label{eq:two_gauss}
P^*(\lambda;\rho,\sigma) = \frac{1}{2 \sigma \sqrt{2 \pi}} \sum_{\epsilon = \pm 1} \exp \left[ - \frac{(\lambda + \epsilon \rho)^2}{2 \sigma^2} \right] .
\end{equation}

\noindent This distribution is a generalization of the one studied by
Watkin and Nadal~\cite{WN1}, who considered optimal learning for
clusters with $\sigma = 1$. If $\rho = 0$, (\ref{eq:two_gauss})
corresponds to the single gaussian scenario studied by Reimann {\it
et al.}~\cite{RV1}. In this paper we investigate the complete phase
diagram in the plane $\rho, \sigma$.

The first two moments of (\ref{eq:two_gauss}) are

\begin{eqnarray}
\langle\lambda \rangle & = & \, 0\\
\langle \lambda^2 \rangle & = & \, \rho^2 + \sigma^2.
\end{eqnarray}

\noindent Thus, if $\sigma = 1$ only distributions with $\langle
\lambda^2 \rangle >1$ are considered. The optimal solution in that
case is close to the one obtained with a quadratic
potential~\cite{WN1}. Quadratic potentials detect the direction
extremizing the variance of the training set, which we call {\it
variance learning}. We show below that the optimal overlap may be
much larger than the one obtained through variance learning if the
clusters have $\sigma < 1$.

\begin{figure}
\centerline{\psfig{figure=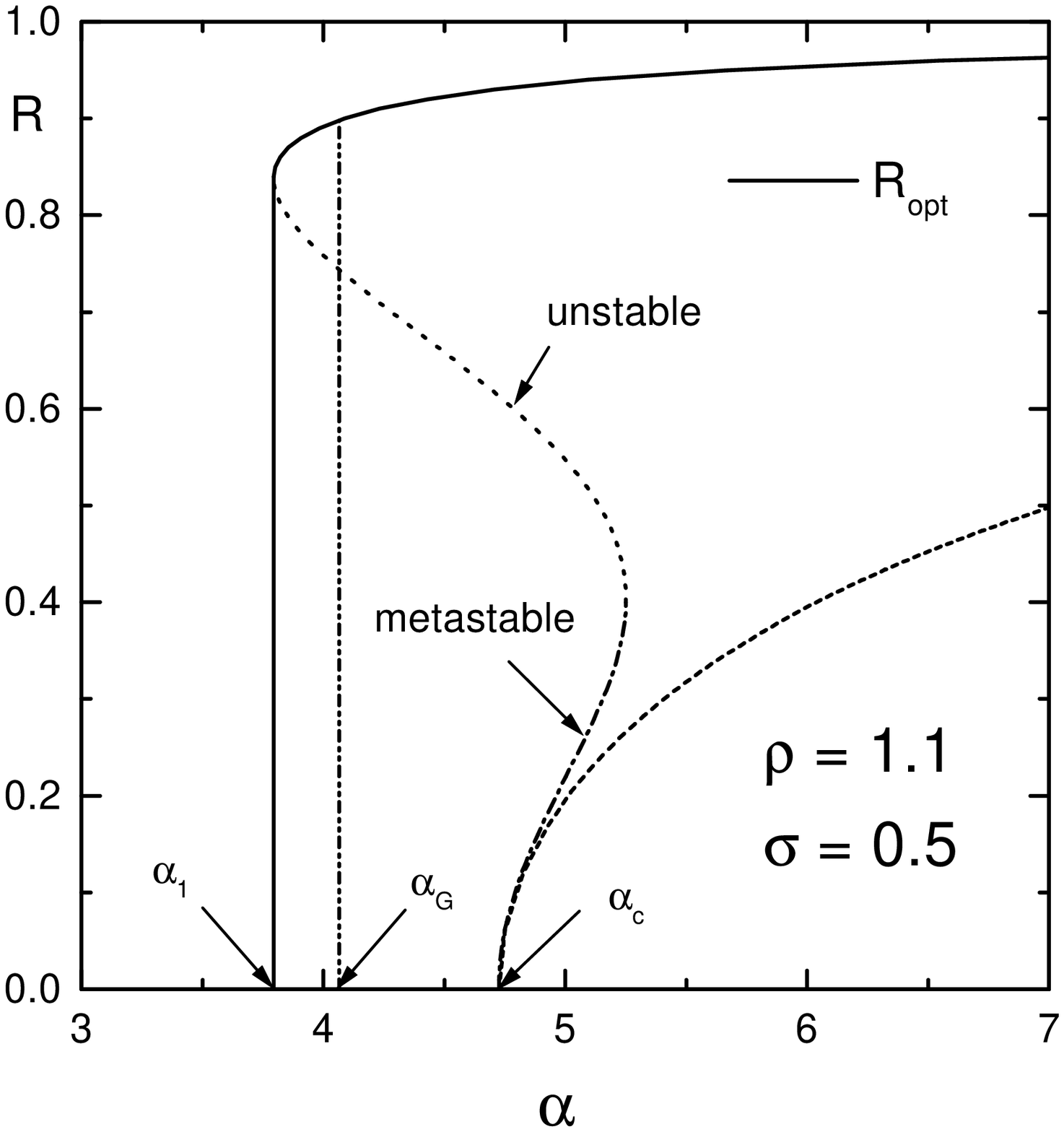,height= 7 cm}}
\caption{Learning curves for the two-clusters scenario, for cluster parameters corresponding to the lowest small square of fig. {\protect \ref{phases}}. Full line: optimal learning. Dash-dotted lines: lower branch of metastable solutions to optimal learning. Also shown: the replica-symmetry unstable curve (dotted line). The lowest dashed line corresponds to learning with a quadratic potential (variance-learning). Here, $\alpha_1=3.79$, $R_{\rm opt}(\alpha_1)=0.84$; the bayesian first order transition occurs at $\alpha_{\rm G}=4.07$, $R_{\rm opt}(\alpha_{\rm G})=0.9$; the critical $\alpha$ for variance-learning is $\alpha_c=4.73$.}
\label{rho1.1}
\end{figure}

\begin{figure}
\centerline{\psfig{figure=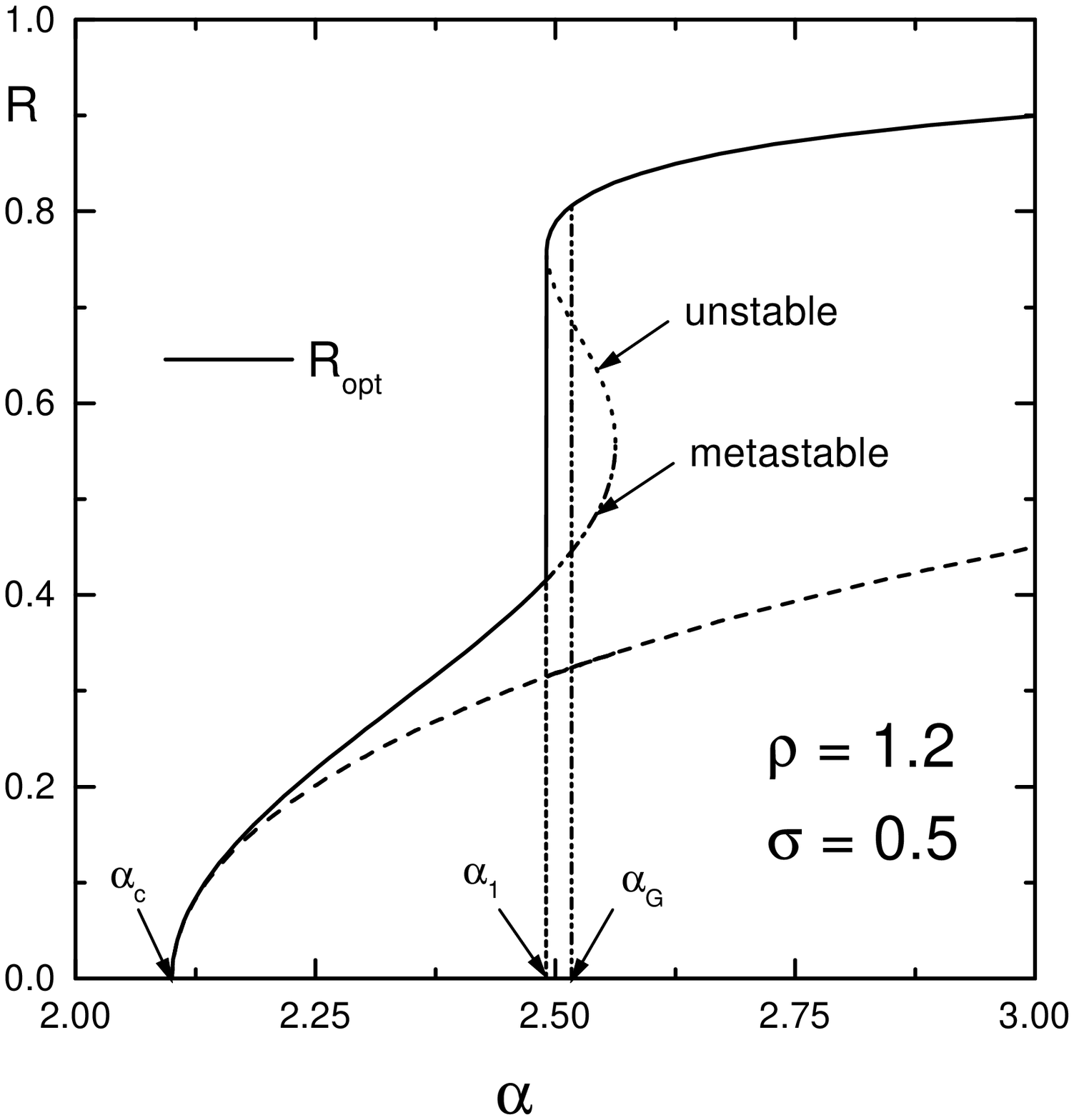,height= 7 cm}}
\caption{Learning curves for the two-clusters scenario, for cluster parameters corresponding to the central small square of fig. {\protect \ref{phases}}. Full line: optimal learning. Dash-dotted lines: lower branch of metastable solutions to optimal learning. Also shown: the replica-symmetry unstable curve (dotted line). The lowest dashed line corresponds to learning with a quadratic potential (variance-learning). Here, $\alpha_1=2.49$, $R_{\rm opt}(\alpha_1)=0.76$; the bayesian first order transition occurs at $\alpha_{\rm G}=2.52$, $R_{\rm opt}(\alpha_{\rm G})=0.81$; the critical $\alpha$ for variance-learning is  $\alpha_c=2.10$.}
\label{rho1.2}
\end{figure}

\begin{figure}
\centerline{\psfig{figure=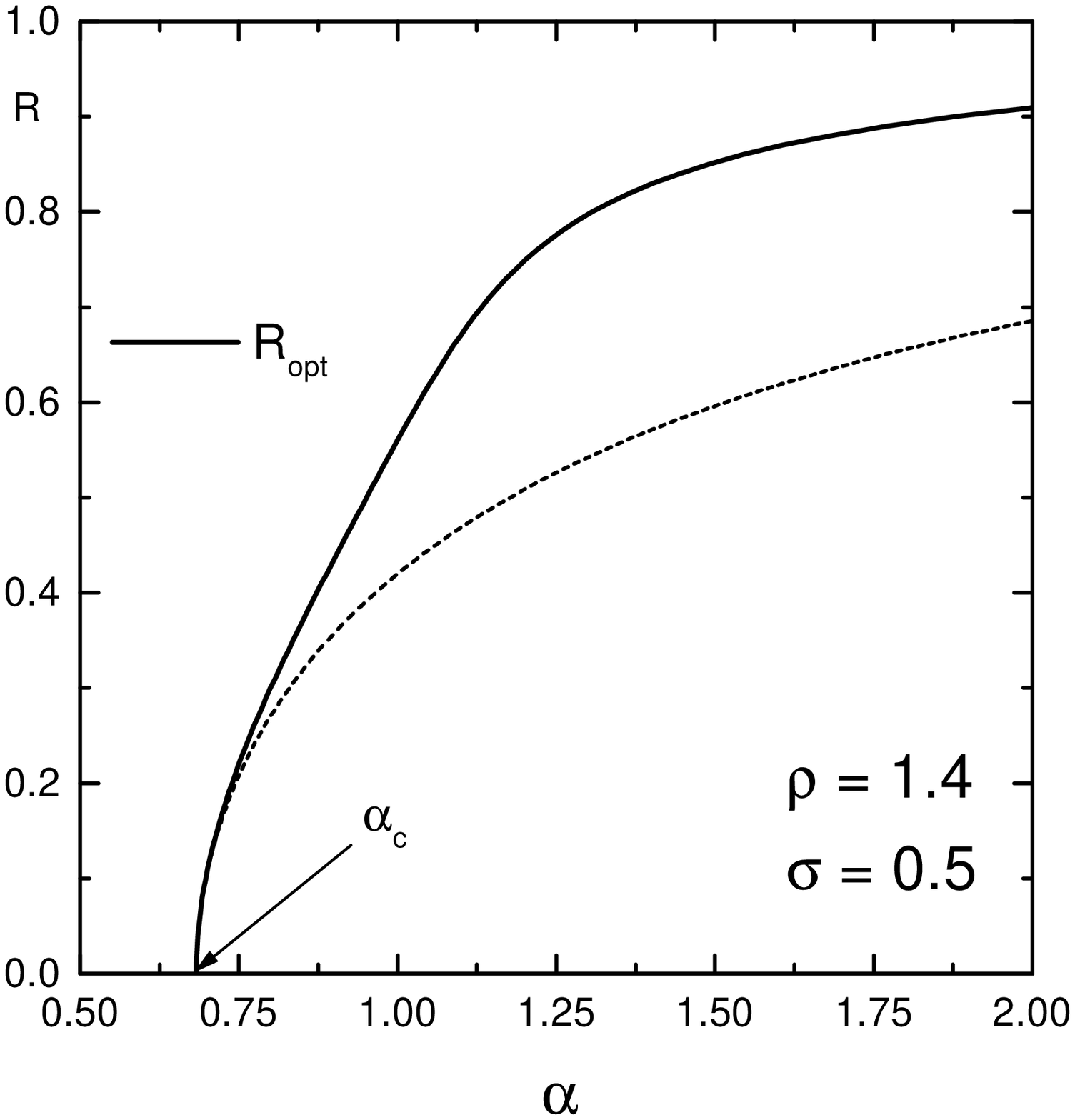,height= 7 cm}}
\caption{Optimal learning curves (full line) for the two-clusters scenario, for cluster parameters corresponding to the upper small square of fig. {\protect \ref{phases}}. The lowest dashed line corresponds to learning with a quadratic potential (variance-learning). Here, $\alpha_c=0.68$.}
\label{rho1.4}
\end{figure}

Introducing the expression of $V^*$ obtained from
(\ref{eq:two_gauss}) and (\ref{eq:broken_proba}) into
(\ref{eq:alpha(R)}) gives $\alpha$ as a function of $R_{\rm opt}$. It
turns out that, for some values of $\alpha$, this function has three
different roots for $R_{\rm opt}(\alpha)$, as is apparent on figures
\ref{rho1.1} and \ref{rho1.2}. The one lying on the branch with
negative slope violates the assumption of replica symmetry. The two
others correspond to minimae of the corresponding free energies.
Figures \ref{rho1.1}, \ref{rho1.2} and \ref{rho1.4} show the optimal
learning curves for several values of $\rho$ and $\sigma$ in the range
not investigated before. The two branches $R_{\rm opt}(\alpha)$ with
positive slope that satisfy condition (\ref{eq:rep_sym}), and the
dotted line of negative slope (inconsistent with the assumption of
replica symmetry), are presented for illustration. 
The value of $\alpha$ at which the jump from one branch to the other occurs is discussed in next section.
The performance
obtained through learning  with simple quadratic potentials is also
presented, to show the dramatic improvement of optimal learning with
respect to variance learning for double clusters with $\sigma <1$.

\section{Bayesian versus optimal solutions.}
\label{B_vs_Opt}

As pointed out in section \ref{Opt_Pot}, eq. (\ref{eq:alpha(R)}) 
may be deduced in two different ways: through the 
determination of the bayesian learning performance, or through 
functional optimization. This procedure yields of a cost function 
for each training set size $\alpha$ whose minimum gives the solution 
with maximal overlap . 

The bayesian solution to the learning problem is given by the 
average of solutions sampled with Gibbs' probability. A simple 
argument~\cite{WN1} shows that the typical bayesian performance 
satisfies $R_{\rm b} = \sqrt{R_{\rm G}}$, where $R_{\rm G}$ is 
the typical overlap between a solution drawn with probability 
(\ref{eq:Gibbs}) and the symmetry-breaking direction ${\bf B}$. 
$R_{\rm G}$ minimizes the free energy with potential 
$V(\gamma)=V^*(\gamma)$ at inverse temperature $\beta = 
1$~\cite{WN1,VR1}. 

As eq. (\ref{eq:alpha(R)}) is satisfied both by $R_{\rm b}$ and 
$R_{\rm opt}$, it is tempting to conclude that bayesian learning 
is optimal. If eq. (\ref{eq:alpha(R)}) has a unique 
solution, this is obviously the case. However, equation (\ref{eq:alpha(R)}) 
may not be invertible. This arises in the two-cluster scenario 
presented in the previous section, where two branches of solutions 
consistent with the assumption of replica symmetry exist for some 
values of $\alpha$. In the case of bayesian learning, these branches 
result from the fact that Gibbs' free energy has two local minima as  
a function of $R$. $R_{\rm G}$, the thermodinamically stable state, 
corresponds to the absolute minimum. When $\alpha$ changes, $R_{\rm G}$ 
jumps from one branch to the other through a first order phase 
transition at $\alpha = \alpha_{\rm G}$, where both minima have the 
same free energy~\cite{Copelli}. Therefore the bayesian solution, 
which is the average of the solutions sampled with Gibbs' probability, 
presents a jump at the same value $\alpha_{\rm G}$ as Gibbs' 
performance. Thus, the metastable states of higher performance 
than $R_{\rm b}$, which exist for $\alpha < \alpha_{\rm G}$, 
cannot be obtained through bayesian learning. 

On the other hand, in section \ref{Opt_Pot} we determined 
optimal potentials whose minimization allow to obtain performance 
$R_{\rm opt}$. These potentials exist for all the pairs 
$(\,\alpha, R_{\rm opt}(\alpha)\,)$ lying on the monotonically 
increasing  branches of $R_{\rm opt}(\alpha)$, which satisfy the 
hypothesis of replica symmetry. Potentials allowing to reach the 
performances of the upper (Gibbs-metastable) branch thus exist. 
It should be noticed that we cannot determine the 
position of the jump of $R_{\rm opt}$ through the comparison of 
the free energies corresponding to solutions on different branches 
at the same $\alpha$, as was done to determine $\alpha_{\rm G}$, 
because a {\it different} potential has to be minimized for each 
pair $(\,\alpha, R_{\rm opt}(\alpha)\,)$ and, as discussed 
in section \ref{Opt_Pot}, these potentials are measured in the 
arbitrary units determined by our choice $c=1$. 

\begin{figure}
\centerline{\psfig{figure=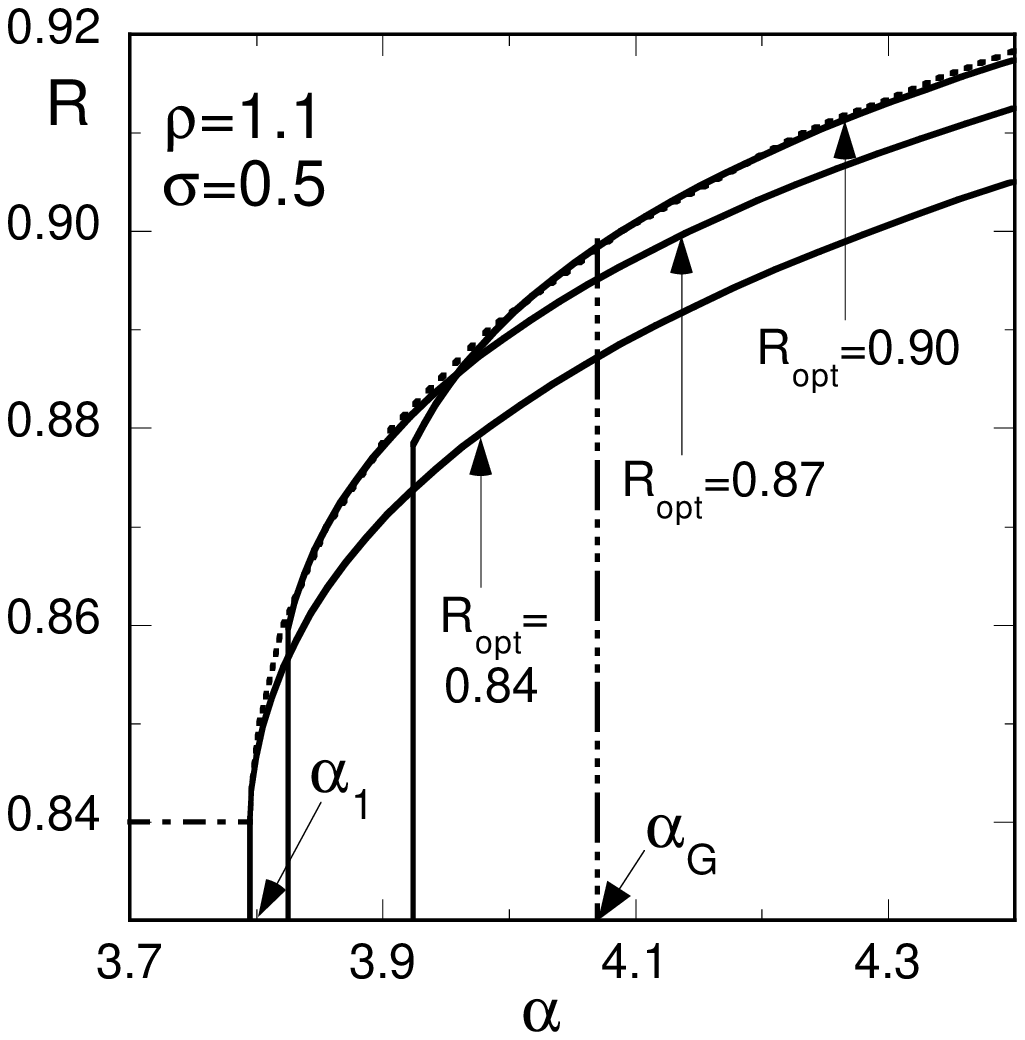,height= 7 cm}}
\caption{Learning curves for $\rho=1.1$, $\sigma=0.5$ obtained with the optimal potentials correponding to $R_{\rm opt}=0.84$, $R_{\rm opt}=0.87$ and $R_{\rm opt}=0.90$ (full lines). Only the solutions consistent with the replica symmetry hypothesis are shown. Dotted lines: optimal solution.} 
\label{fig:R_metast_11}
\end{figure}

\begin{figure}
\centerline{\psfig{figure=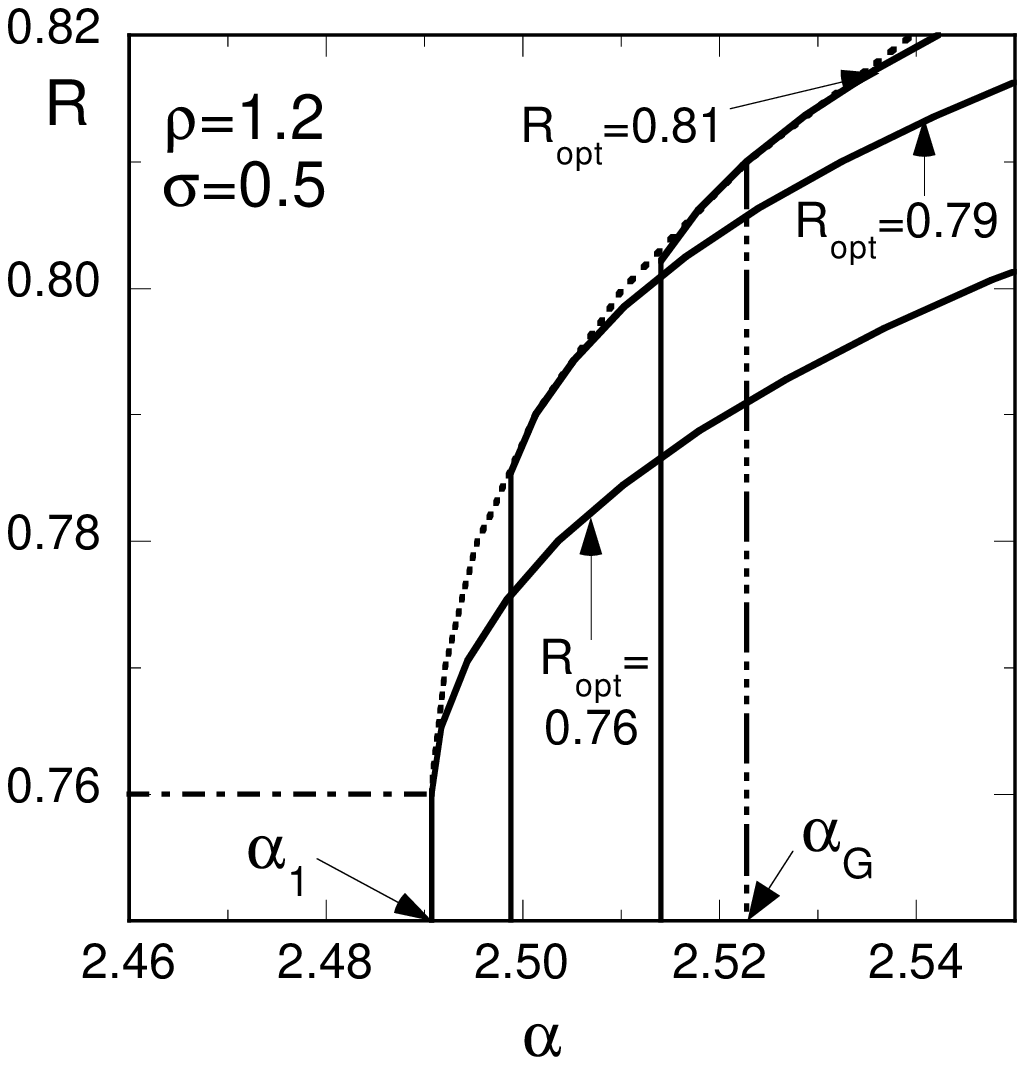,height= 7 cm}}
\caption{Learning curves for $\rho=1.2$, $\sigma=0.5$ obtained with the optimal potentials correponding to $R_{\rm opt}=0.76$, $R_{\rm opt}=0.79$ and $R_{\rm opt}=0.81$ (full lines). Only the solutions consistent with the replica symmetry hypothesis are shown. Dotted lines: optimal solution.} 
\label{fig:R_metast_12}
\end{figure}

In order to clarify this problem, we studied the performance of 
the minima of the optimal potentials. In fact, the properties 
of each of the potentials $V_{\rm opt}(\lambda)$ may be determined 
for any value of $\alpha$ (besides the value for which it has been 
optimized) in the same way as those of other {\it ad hoc} potentials,
by solving numerically eq. (\ref{eq:R1_R2}). Figs. 
\ref{fig:R_metast_11} and \ref{fig:R_metast_12} presents several 
learning curves $R(\alpha)$ obtained with potentials 
$V_{\rm opt}$ optimized for ovarlaps lying on the upper 
metastable branch of Gibbs' learning. They correspond to 
the same clusters' parameters as figs. \ref{rho1.1} and \ref{rho1.2}. 
Each learning curve is tangent to the optimal learning
curve at the point $(\,\alpha(R_{\rm opt}),R_{\rm opt}\,)$ at which 
the potential was determined. This result holds in
particular for all the points lying on the high-performance metastable
branch of bayesian learning, {\it i.e.} for $\alpha_1 < \alpha < 
\alpha_{\rm G}$. It is important to point out that the free energy 
(\ref{eq:f(R,c)}) presents a {\it unique} replica symmetric minimum 
as a function of $R$ for all these potentials. Thus, these results 
show that the corresponding optimal potentials $V_{\rm opt}$ allow to 
select, among the metastable states of Gibbs learning, the one of 
largest overlap. In particular, the Gibbs' metastable states in 
the upper branch for $\alpha < \alpha_{\rm G}$ are learnable through the minimization of the corresponding optimal potential. Thus, in the range 
$\alpha_1 < \alpha < \alpha_{\rm G}$ bayesian learning is not optimal. 
This surprising behavior may arise whenever the curve $R_{\rm G}(\alpha)$ 
of Gibbs learning presents first order phase transitions.

It is worth noting that, besides the solutions that verify 
the replica symmetric condition (\ref{eq:rep_sym}), solutions unstable 
under replica symmetry breaking with smaller $R$ and slightly higher 
free energy also exist. The nature of these states is
very different from that of the metastable states of Gibbs
learning. Whether the typical performance in the case of the double
cluster distributions is the one described by the replica
symmetric solution or not remains an open problem.

\section{The phase diagram}
\label{sec:phase_diag}

\begin{figure}
\centerline{\psfig{figure=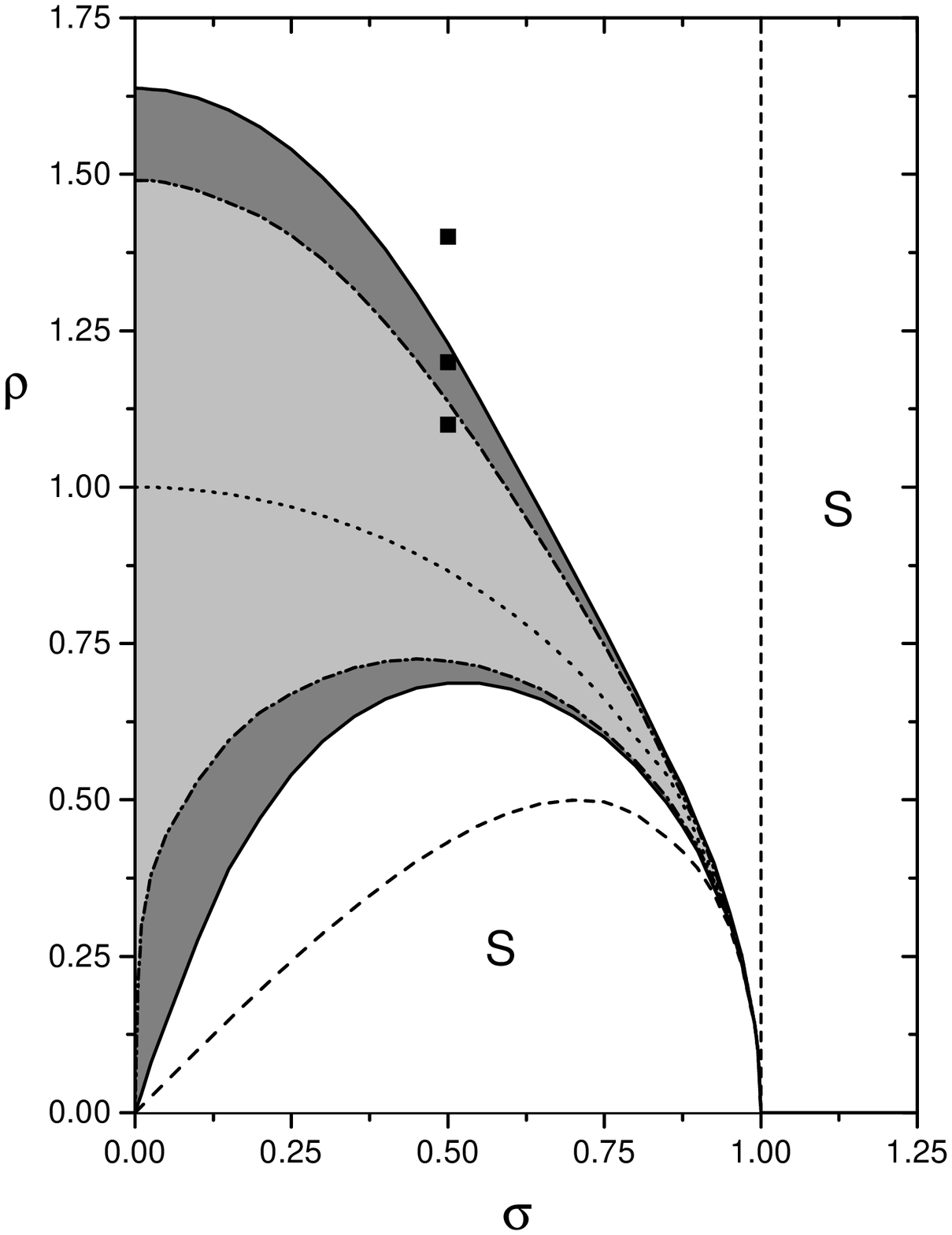,height= 12 cm}}
\caption{Phase diagram of the two-cluster scenario. The three small squares correspond to the learning curves of figs. {\protect \ref{rho1.1}}, {\protect \ref{rho1.2}} and {\protect \ref{rho1.4}}.}
\label{phases}
\end{figure}

In this section we describe, on the $\rho-\sigma$ plane, all the
possible learning phases that may arise in unsupervised learning
within the two gaussian-clusters scenario. As shown on fig.
\ref{phases}, depending on the values of $\rho$ and $\sigma$,
qualitatively different behaviours of the learning curves $R_{\rm
opt}(\alpha)$ may appear. They are correlated with the form of the
corresponding optimal potentials.

The regions marked with an "S" are regions of variance-type learning:
the optimal potential is a single well with $V_{\rm opt} \rightarrow
+\infty$ for $\lambda \rightarrow \pm\infty$ if $\sigma^2 < 1$, and
$V_{\rm opt} \rightarrow -\infty$ for $\lambda \rightarrow \pm\infty$
if $\sigma^2 > 1$. In these regions, the learning curves increase
monotonically with $\alpha$, starting at $\alpha_c=|\langle \lambda^2
\rangle-1|^{-2}$, like for quadratic potentials~\cite{RV1}.

\begin{figure}
\centerline{\psfig{figure=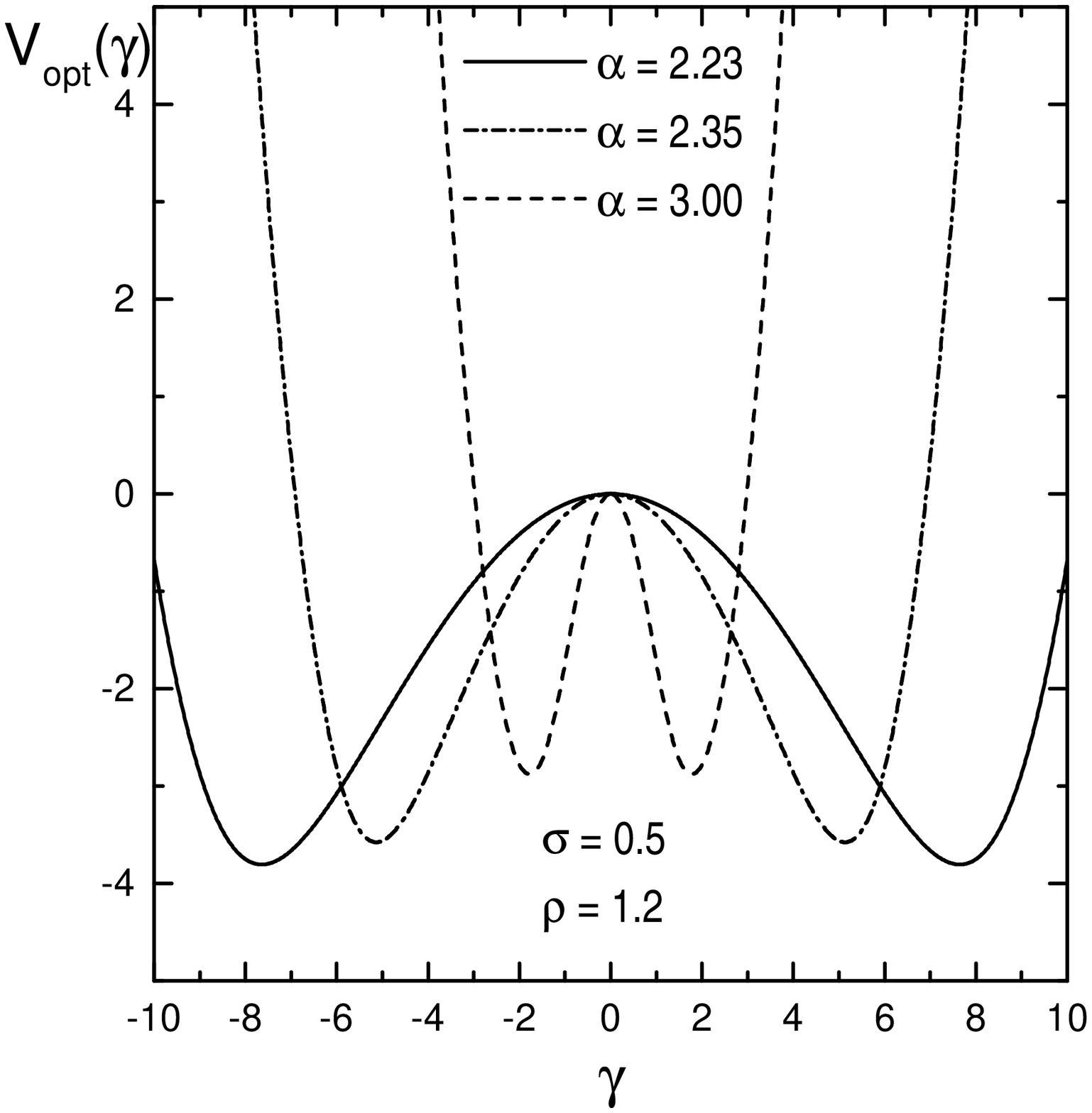,height= 7 cm}}
\caption{Potentials for optimal learning in the grey regions of the phase diagram, showing the evolution of the separation between minima with $\alpha$.}
\label{potentiel}
\end{figure}

For parameter values outside the "S" regions, $V_{\rm opt}
\rightarrow +\infty$ for $\lambda \rightarrow \pm\infty$, even in
the large variance region $\langle \lambda^{2} \rangle > 1$ where
naively one would expect the potential to have the same asymptotic
behaviour as for $\sigma^2 > 1$. Depending on the value
of $R_{\rm opt}$, the optimal potential may be a double-well function
of the local field $\gamma$. In the latter case, the optimal
learning strategy looks for structure in the data distribution
rather than for directions extremizing the variance. This is more
striking on the line $\langle \lambda^2 \rangle = 1$ corresponding to
distributions with the same second moment in all the directions. On
this line, variance learning is impossible and $\alpha_c= \infty$.
However, in the entire light-grey region including this line,
performant learning is achieved if the adequate potential is
minimized. The optimal overlap presents jumps from $R_{\rm opt}=0$ to
finite-$R$ at a fraction of examples $\alpha < \alpha_c$. In the
high-performance branch, the optimal potential is double-well, with
the two minima close to $\pm \rho$, as shown on figure
\ref{potentiel}. Thus, the potential is sensitive to the two
cluster structure, and its minimization results in high performance
learning. For $\rho$ and $\sigma$ in the dark-grey regions, a first
order transition to large $R$ also takes place, but for $\alpha >
\alpha_c$. Below the transition, optimal learning is mainly
controlled by the variance of the training set. 

\begin{figure}
\centerline{\psfig{figure=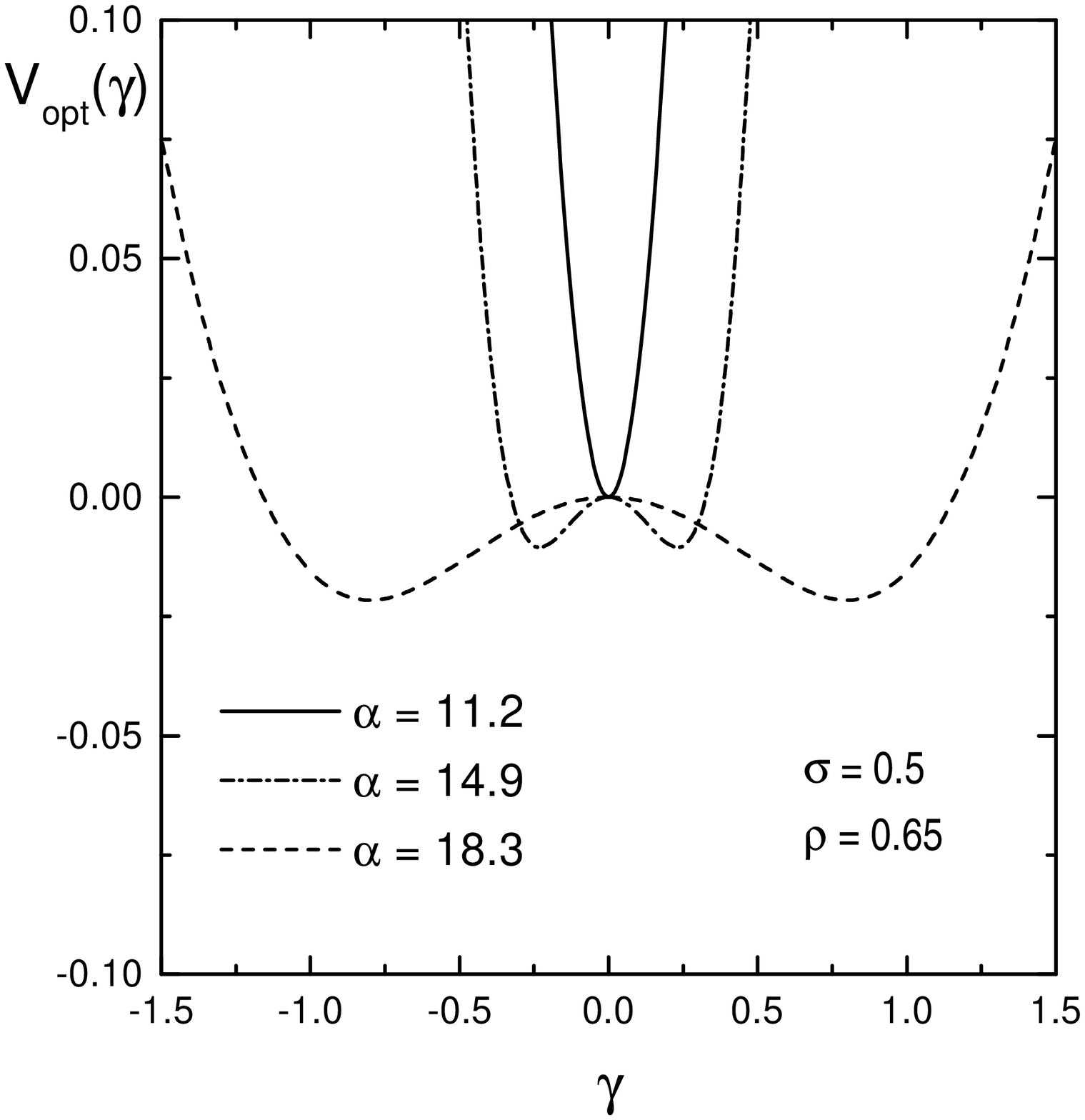,height= 7 cm}}
\caption{Potentials for optimal learning in the white regions of the phase diagram, showing the appearence of the two minima that get farther apart with increasing $\alpha$.}
\label{potentiel2}
\end{figure}

In the white regions on both sides of the dark-grey ones, no first
order phase transitions to high performance learning occur as a
function of $\alpha$. In the white region just below the dark-grey
one, the potential changes smoothly from a single to a double well
with increasing $R_{\rm opt}$. The two minimae appear at $\gamma=0$,
and move away with increasing $R_{\rm opt}$, as shown on fig.
\ref{potentiel2}. However, as far as these minimae are not
sufficiently apart, $R_{\rm opt}$ remains close to the values
obtained with simple quadratic potentials. Conversely, in the upper
white region, which corresponds to $\langle \lambda^2 \rangle \gg 1$
the minima of the optimal potential are far appart, in a region of
large local fields, where the patterns' distribution is vanishingly
small. Thus, in the range of pertinent values of $\gamma$ the
potential is concave ($V_{\rm opt}^{''} < 0$), and here also, like in
the lower white region, the values of $R_{\rm opt}$ are close to
those obtained with quadratic potentials~\cite{RV1}.
 
\section{Conclusion}
\label{sec:conclusion}

Learning the symmetry-breaking direction of a distribution of
patterns with axial symmetry in high dimensions is a difficult
problem. In this paper we determined the optimal
performances that may be reached if the patterns distribution has a
double-cluster structure in the symmetry-breaking direction.
Depending on the clusters' size and separation, the learning curves
may present several phases with increasing $\alpha$, including
novel first order transitions from low-performance variance learning
to high-performance structure detection. We showed that when the
optimal learning curves present such discontinuities, bayesian
learning may be not optimal. These results rely on the assumption
that the solution with replica symmetry is the absolute minimum of
the free energies studied. Although we showed that our solutions
satisfy the replica symmetry stability condition, we cannot rule out
the existence of states of lower energy, but having broken replica
symmetry.

\subsection*{Acknowledgements}
We thank Chris Van den Broeck and Mauro Copelli for helpful discussions at the early stages of this work.

\end{document}